\documentclass[fleqn,12pt,twoside]{article}
\usepackage{espcrc1}
\usepackage{epsf,epsfig}

\usepackage{graphicx}
\usepackage[figuresright]{rotating}

\newcommand{\AmS}{{\protect\the\textfont2
  A\kern-.1667em\lower.5ex\hbox{M}\kern-.125emS}}

\title{ {\bf $\phi$ meson production in Au - Au collisions at $\sqrt{s_{NN}}$ = 200 GeV 
       }}

\author{Debsankar Mukhopadhyay \address[WIS]{Particle Physics Department, 
        Weizmann Institute of Science, 
        Rehovot 76100, Israel} for the PHENIX Collaboration {\thanks{ for the full PHENIX Collaboration author list and acknowledgements, see Appendix ``Collaborations" of this volume.}}}%
       
\begin{document}

\maketitle

\begin{abstract}

A simultaneous  measurement of the
$\phi$ meson via its $K^{+} K^{-} $ 
and $e^{+} e^{-}$ decay channels was performed in Au + Au collisions
at $\sqrt{s_{NN}}$ = 200 GeV at mid-rapidity by the PHENIX experiment. 
The preliminary minimum bias yields $dN_{\phi}/dy$  
in the kaon and electron channels are 
$2.01  \pm  0.22 (stat.) ^{+1.01}_{-0.52} (syst.)$ and
$5.4 \pm 2.5 (stat) ^{+3.4}_{-2.8} (syst.)$, respectively.
The  centrality dependence of the yield in the $K^{+} K^{-} $ channel is
presented.
\end{abstract}

\section{Introduction}

The $\phi$ meson is an important probe for studying relativistic heavy
ion collisions. 
Since the mass of the $\phi$ meson is close to
twice the kaon mass, any medium modification of its spectral shape (mass and/or width) {\cite{rapp}} as chiral symmetry restoration is approached may induce a change
in its branching ratio in the kaon channel. 
The simultaneous measurement of $\phi$ decay  into $K^+K^-$
and $e^+e^-$ is a very powerful tool in the search for such in-medium modifications.
Consisting of $s \bar{s}$, the $\phi$ meson is also a sensitive probe of strangeness production {\cite{rafelski}}. 
The PHENIX experiment
at RHIC, with its excellent mass resolution and particle identification capability, (comparable to or better
than the natural width of the $\phi$ meson)  has the unique capability to
measure the $\phi$ meson through both $K^{+} K^{-}$ and $e^{+} e^{-}$
decay channels at mid-rapidity. We present the preliminary results of 
the $\phi$ meson 
measurement made during the 2001 RHIC run.

\section{Experimental setup and data analysis}

The results presented here were obtained using the two central arms of the PHENIX
spectrometer {\cite{tat}}. 
The $\phi \rightarrow K^{+} K^{-}$ analysis used the drift chamber (DC), 
two sets
of multiwire proportional chambers with pixel-pad readout (PC1 and PC3), and the
time of flight (TOF) module of the east arm. The acceptance is defined by the TOF module which covers  the pseudorapidity range 
$|\eta| < 0.35$ and an azimuthal range of $\Delta\phi \approx 30^{\circ}$.
The kaons were identified via reconstructed momentum combined with a TOF measurement
with a time resolution $\sim 120$ ps.
With a 2$\sigma$ momentum dependent
cut in the mass squared distributions, $\pi/K$ were well separated up to $p \sim 2.0$ GeV/c. 

 The $\phi \rightarrow e^{+} e^{-}$
measurements were performed with the 
DC, PC1, PC3, the Ring Imaging Cerenkov Detector
(RICH) and the electromagnetic calorimeter (EMCAL) of both the east and west arm, each one
covering 
$\Delta\phi = 90^{\circ}$ and $|\eta| < 0.35$. 
Electrons were identified primarily by the RICH. Further identification was provided by   requiring
the energy in the EMCAL  to match the measured momentum of the tracks.

Charged particle tracks and their momenta
were reconstructed using the DC and PC1. Tracks were
confirmed by a matching hit in  
PC3 and TOF in the case of kaons, and in  the EMCAL for electrons.

The beam-beam counters
(BBC) and the zero-degree calorimeters (ZDC) provided the trigger and  were used to
determine the event  centrality. The BBC were also used to determine the z-coordinate 
of the collision vertex ($z_{vertex}$). 
The analysis used $27.4 \times 10^{6}$ (for kaons) and 
$25.8 \times 10^{6}$ (for electrons) minimum bias events with a vertex position within $|z_{vertex}| < 30$ cm.

\section{Analysis procedure}

\subsection{Data}

All identified kaon tracks   in a given event were combined to form
the invariant mass distributions of  the like sign ($N_{++}$, $N_{--}$) and unlike sign
pairs. The large combinatorial background inherent to this procedure 
was estimated
by an event mixing method in which all $K^{+}$ tracks from one event
were combined with $K^{-}$ tracks of ten other events 
of  the same centrality and vertex class. The mixed event invariant
mass distribution was then normalized to the measured $2\sqrt{N_{++}N_{--}}$. 
The validity of the method was tested by constructing, in a similar way, a
combinatorial like-sign spectrum and comparing it to the measured like-sign pair 
distribution. 
Finally, the signal
was obtained by subtracting the mixed event spectrum  from the measured one. 
This gives the
uncorrected  $\phi$-yield ($N^{Signal}_{\phi}$) of the $K^{+}K^{-}$ 
decay channel. An identical procedure was used for the measurement
of the $e^{+}e^{-}$ decay channel.

\subsection{Corrections}

In order to obtain the yield $dN_{\phi}/dy$ from the uncorrected signal, 
we correct our invariant
mass distributions for detector acceptance and reconstruction efficiency.
The correction was determined using
Monte Carlo simulations in two steps. In the first step, we generated
single $\phi$ mesons with an exponential transverse momentum distribution 
($dN/dp_{T} \sim p_{T} ~.~ exp(-m_{T}/T)$),  assuming a temperature of {\it T} = 380 MeV
 which is consistent with  the measurement by the STAR experiment 
at $\sqrt{s_{NN}} = 130$ GeV {\cite{star}}. The generated $\phi$'s were 
then propagated
through the PHENIX detector simulation and the  pair acceptance
and reconstruction efficiency 
was calculated by:
\begin{equation}
\epsilon = \frac{N^{generated}_{\phi}}{N^{reconstructed}_{\phi}}
\end{equation}
In the second step, the multiplicity dependent efficiencies ($\epsilon_{mult}$) 
were estimated 
by embedding single tracks into the real data events.

The final corrected yield is given by
\begin{equation}
\frac{dN_{\phi}}{dy} = \frac{{N^{Signal}_{\phi}} \times \epsilon \times \epsilon^{2}_{mult}}{N_{event} \times BR},
\end{equation}
where BR represents the branching ratio for the specific decay channel 
($e^{\pm}$ or $K^{\pm}$).

\section{Results}

\subsection{Invariant mass distributions}

Fig. 1 shows the minimum bias invariant mass distributions for 
$\phi \rightarrow K^{+} K^{-}$ (panel (a)) and $\phi \rightarrow e^{+} e^{-}$ 
(panel (c)). 
The kaon channel results for the 40\%-80\%
centrality range are shown in Fig. 1(b), where we see a considerable improvement
in the signal to background ratio compared with minimum bias spectrum.
Since the combinatorial background increases quadratically with  centrality,
this means that the signal
increases at a much lower rate with centrality.

\begin{tabular}{lll}
\vspace{-0.7cm} \centering
\epsfig{file=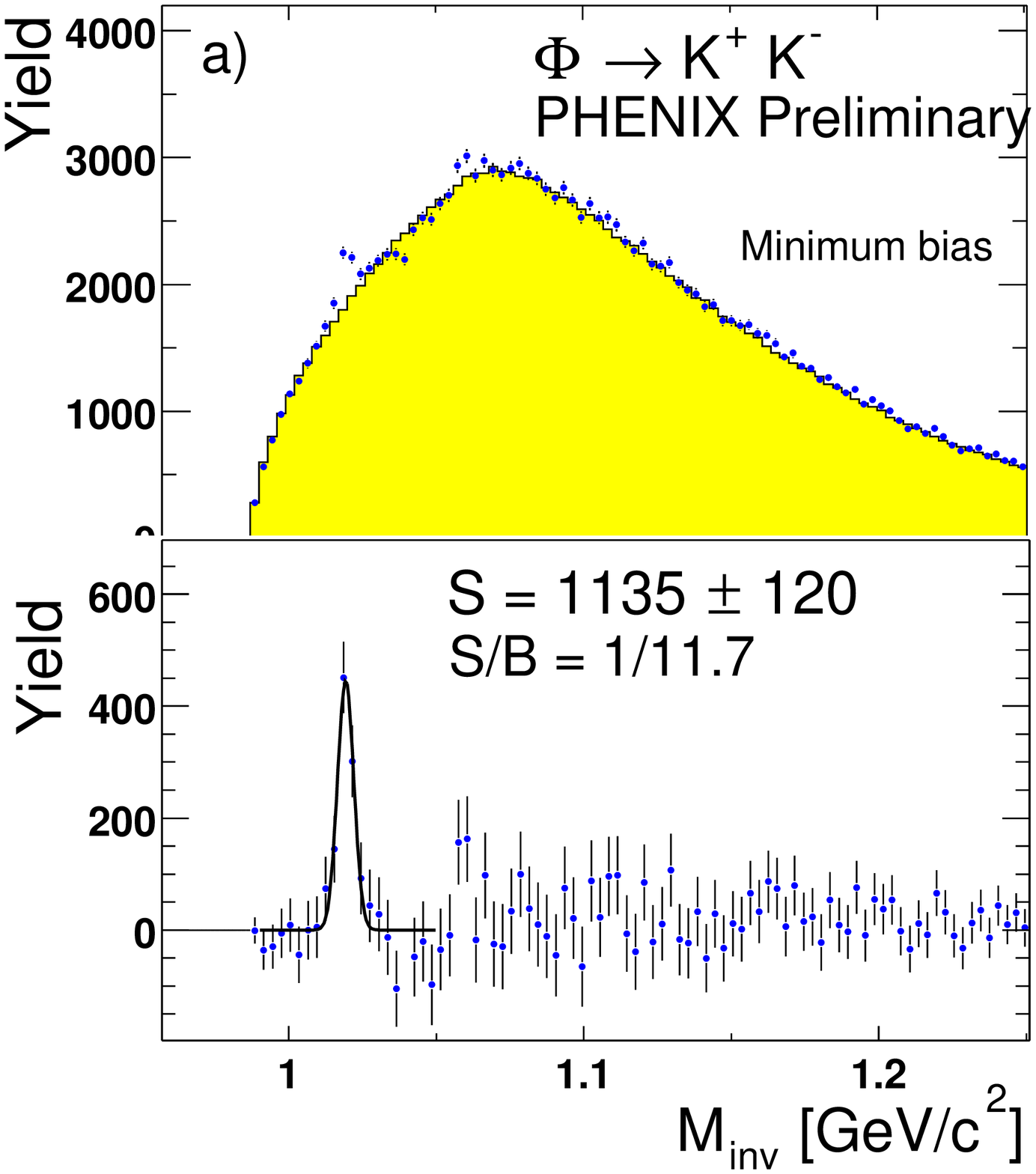,width=5cm,height=6.0cm}&\epsfig{file=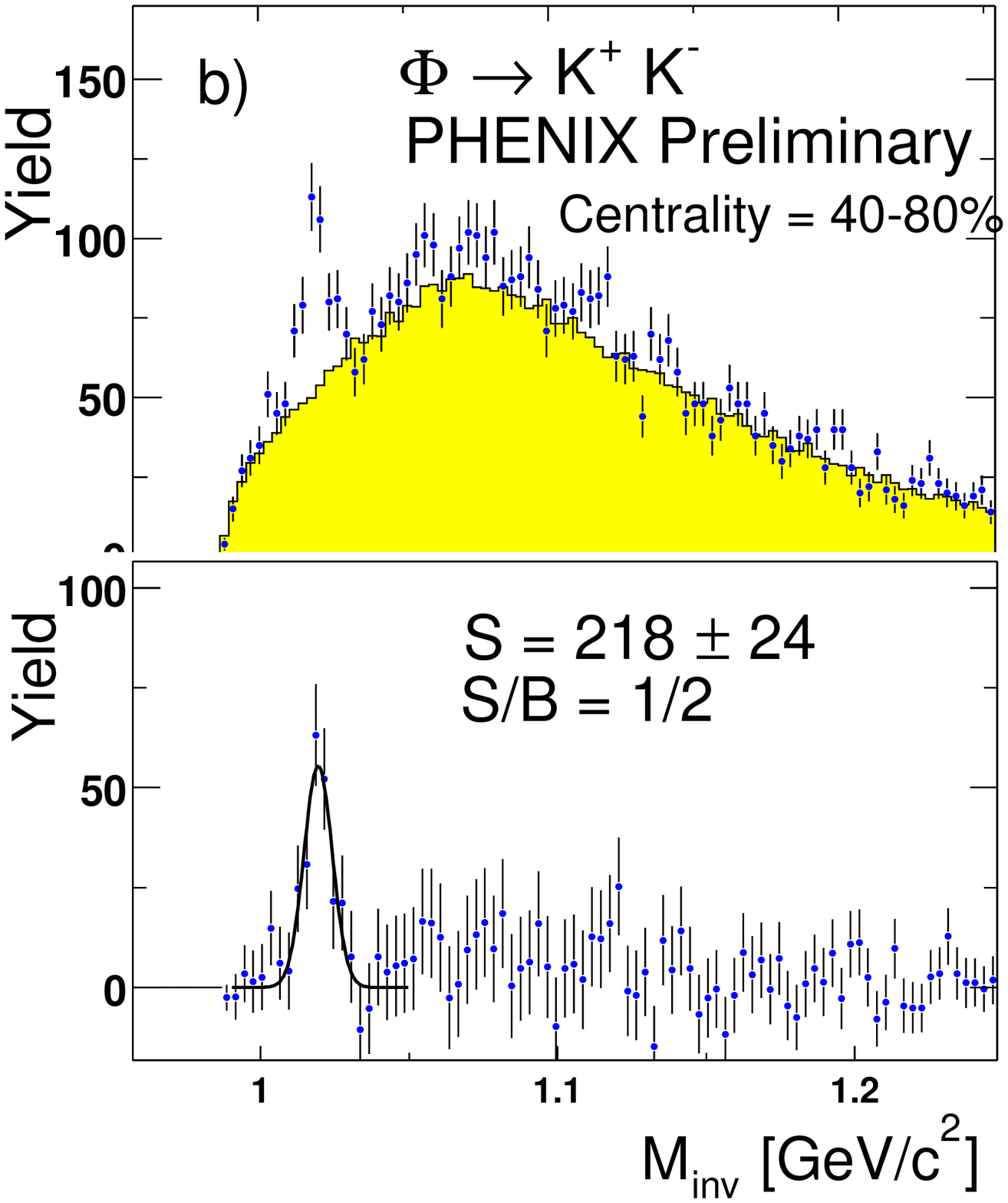,width=5cm,height=6.0cm}&\epsfig{file=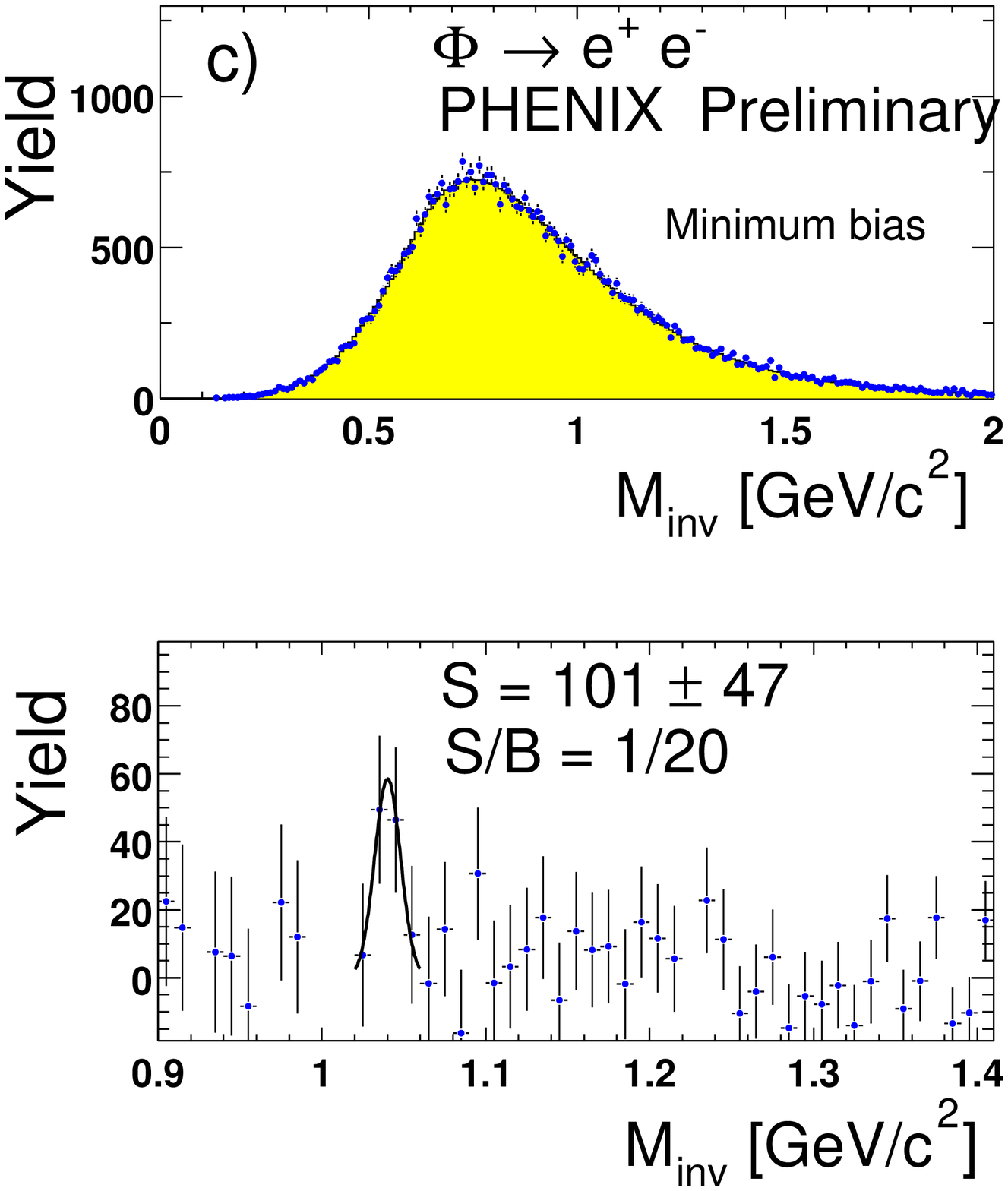,width=5cm,height=5.9cm}
\end{tabular}

\vskip 0.8cm
Figure 1: Invariant mass distributions of $\phi \rightarrow K^{+} K^{-}$ for (a) minimum bias events and 
b) 40-80\% central events. 
The mass spectrum of $\phi \rightarrow e^{+} e^{-}$ for minimum bias is shown
in (c). The points in
the upper panels represent the data and the filled histograms show the combinatorial 
background whereas the lower panels exhibit the subtracted mass spectra.


\subsection{$dN_{\phi}/dy$}

The rapidity density ($dN_{\phi}/dy$) values calculated using Eq. (2)  
for the kaon and
electron channels are shown in Table 1. The systematic errors in both cases originate
from the correction factors in the Monte Carlo simulation and
the analysis method.
In the
$\phi \rightarrow K^{+}K^{-}$ analysis, the main systematic uncertainty  
originates from the uncertainty (assumed to be $\pm 20\%$) in the  temperature used in  the Monte Carlo calculations.
Since the TOF array has no acceptance for low pair $p_{T}$, 
the correction factor $\epsilon$ is very sensitive to the slope of 
the parent $p_{T}$ distribution.
For $\phi \rightarrow e^{+}e^{-}$, the
analysis procedure is the major source of the systematic error. Within the present large statistical and systematic errors, the values of
$dN_{\phi}/dy$ for both cases are consistent with each other.
$\phi$ mesons
from dimuon and kaon channels were also observed  by the fixed target 
NA50 and NA49 experiments respectively at a beam energy of 158 AGeV at CERN.
These two independent experimental results showed a factor of 5 difference in the 
$dN_{\phi}/dy$ values for
the muon channel ($dN_{\phi}/dy \sim 13$) as compared to the kaon channel 
($dN_{\phi}/dy \sim 2.35$) {\cite{ror}}. Higher statistics together with 
a better understanding of the systematics will enable us to see whether or not
such a large difference persists at
RHIC energies.

\begin{table}
\vspace{-0.5cm}
\centering
Table 1: $\phi$ - meson rapidity density in Au - Au at $\sqrt{s_{NN}}$ = 200 GeV
\vskip 0.1cm
\begin{tabular}{|l|c|c|c|}\hline
Decay channel&$dN_{\phi}/dy$&Statistical error&Systematic error\\\hline\hline
&&&\\
$\phi \rightarrow K^{+}K^{-}$&2.01&$\pm0.22$&{\large $^{+1.01}_{-0.52}$}\\
&&&\\
$\phi \rightarrow e^{+}e^{-}$&5.4&$\pm2.5$&{\large $^{+3.4}_{-2.8}$}\\
&&&\\\hline
\end{tabular}
\end{table}

\subsection{Centrality dependence}

In Fig. 2(a),  we plot  $dN_{\phi}/dy$ as a function of centrality (expressed in
percentiles)  for the kaon channel. 
The variation of the rapidity density per participant 
with the number of participants is  shown in Fig. 2(b).
The shape of this curve differs
from the  $(1/N_{part}) \times (dN_{ch}/d\eta$) vs. $N_{part}$ distribution, which
shows a steady increase of $\sim 20\%$  from  peripheral to central collisions {\cite{sura}}.

\begin{tabular}{ll}
\vspace{-0.8cm} \centering
\epsfig{file=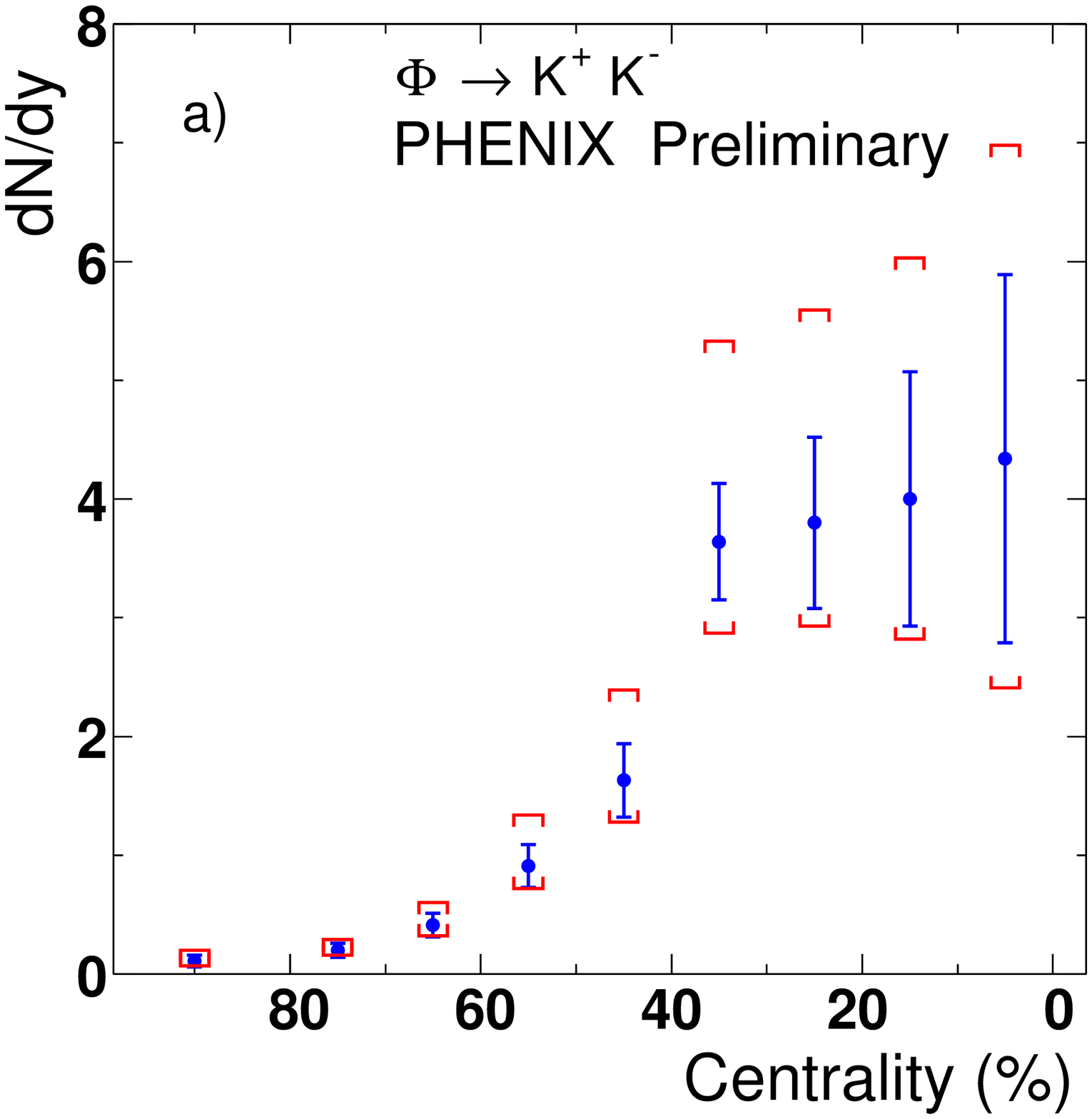,width=6.0cm,height=5.8cm}&\epsfig{file=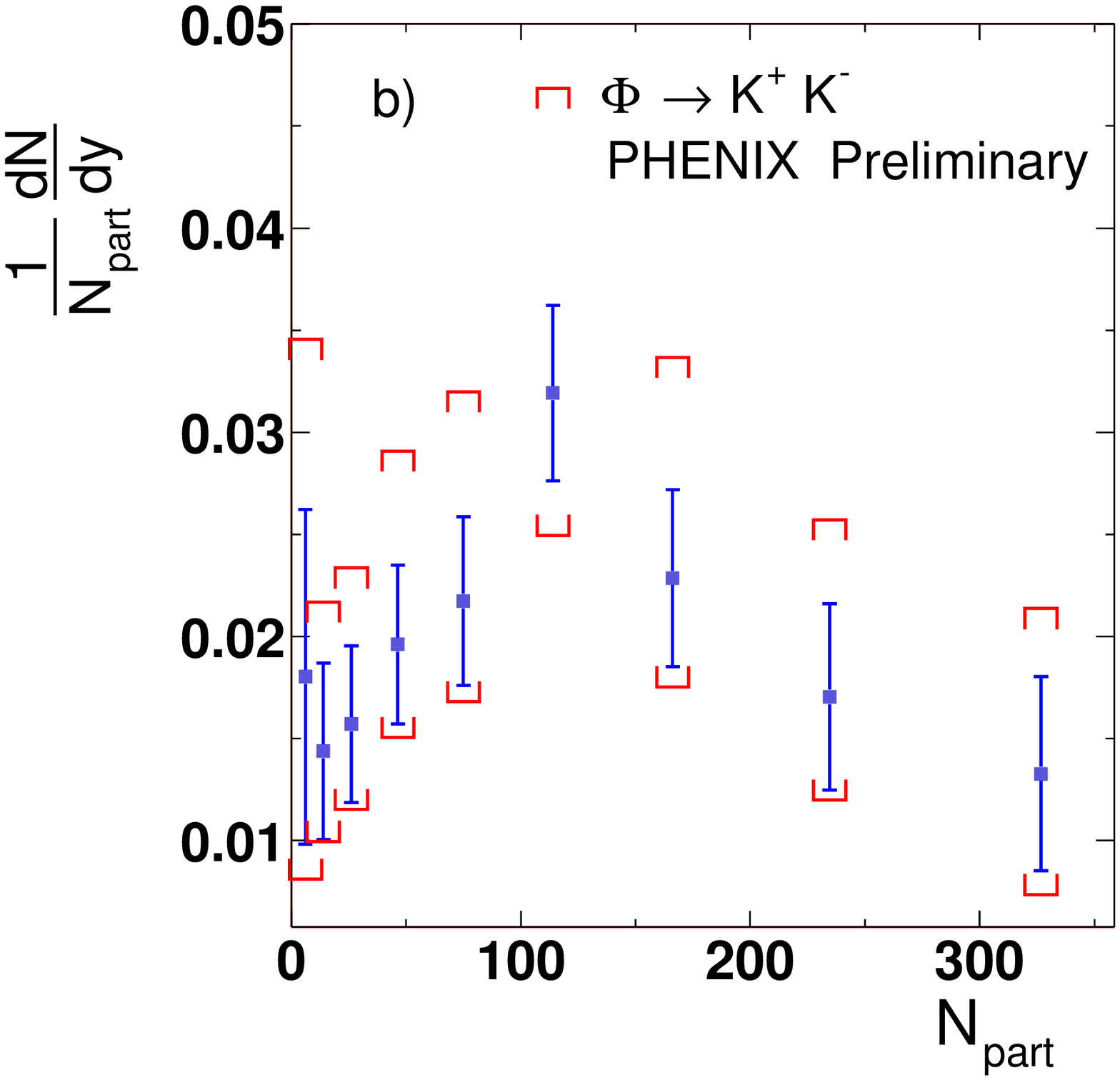,width=7.20cm,height=5.8cm}
\end{tabular}
\vskip 0.8cm

Figure 2: Centrality dependence of $\phi$ yields (solid lines are statistical and brackets represent systematic errors).

\section{Summary}

PHENIX has measured the yield of $\phi$ mesons via the $K^{+} K^{-}$ and $e^{+} e^{-}$ channels.
The values of $dN_{\phi}/dy$ for $\phi \rightarrow K^{+}K^{-}$ and $\phi \rightarrow e^{+}e^{-}$ are
 consistent within the large statistical and systematic  errors of the present
preliminary analysis.
The centrality dependence 
of the $\phi$ yields in the kaon decay channel is also presented.

\end{document}